\documentclass[aps,prl,reprint,superscriptaddress]{revtex4-1}

\usepackage{graphicx}
\usepackage{siunitx}
\usepackage{overpic}
\usepackage[colorlinks=true,citecolor=blue, linkcolor=blue, urlcolor=blue]{hyperref}

\begin{document}

\preprint{v0p8}

\title{Interaction-assisted quantum tunneling of a Bose-Einstein condensate out of a
single trapping well}


\author{Shreyas \surname{Potnis}}
\email[]{spotnis@physics.utoronto.ca}

\author{Ramon \surname{Ramos}}
\affiliation{Centre for Quantum Information and Quantum Control and Institute for Optical Sciences, Department of Physics, University of Toronto, 60 St. George Street, Toronto, Ontario M5S 1A7, Canada}

\author{Kenji \surname{Maeda}}
\author{Lincoln D. \surname{Carr}}
\affiliation{Department of Physics, Colorado School of Mines, Golden, Colorado 80401, USA}

\author{Aephraim M. \surname{Steinberg}}
\affiliation{Centre for Quantum Information and Quantum Control and Institute for Optical Sciences, Department of Physics, University of Toronto, 60 St. George Street, Toronto, Ontario M5S 1A7, Canada}
\affiliation{Canadian Institute For Advanced Research, 180 Dundas Street West, Toronto, Ontario M5G 1Z8, Canada}



\date{\today}

\begin{abstract}

We experimentally study tunneling of Bose-condensed $^{87}$Rb atoms
prepared in a quasi-bound state and observe a non-exponential decay caused by
interatomic interactions. A combination of a
magnetic quadrupole trap and a thin \SI{1.3}{\micro m}  barrier created using a
blue-detuned sheet of light is used to tailor traps with controllable depth
and tunneling rate. The escape dynamics strongly depend on the
mean-field energy, which gives rise to three distinct regimes--- classical
spilling over the barrier, quantum tunneling, and decay dominated by
background losses. We show that the tunneling rate depends exponentially
on the chemical potential. Our results show good agreement with
numerical solutions of the 3D Gross-Pitaevskii equation.
\end{abstract}

\pacs{}

\maketitle


The escape of a particle due to tunneling from a quasibound state is one
of the earliest problems studied in quantum mechanics. When applied to
understand $\alpha$-decay of nuclei, it
successfully explained not only the random nature of the decay,
but also the large range of nuclear lifetimes, spanning many orders
of magnitude~\cite{Gurney1929}. Since then,
quasibound states and quantum tunneling have been shown to play key roles in
physical chemistry~\cite{Gatteschi2003},
biology~\cite{Collini2010}, and condensed matter physics~\cite{Choi2000}.
Consequently, tunneling has been studied in numerous systems, but mainly in
those where the decay is irreversible and
particles decay independently of one another, leading to the usual
exponential dependence of the survival probability with time.
In contrast to all these contexts, in this Letter we report non-exponential
decay of a Bose-Einstein Condensate(BEC) from a quasibound state,
arising due to interatomic interactions.

\begin{figure}[!hb]
    \centering
    \begin{minipage}{.5\columnwidth}
        \centering
        \begin{overpic}[width=\columnwidth]{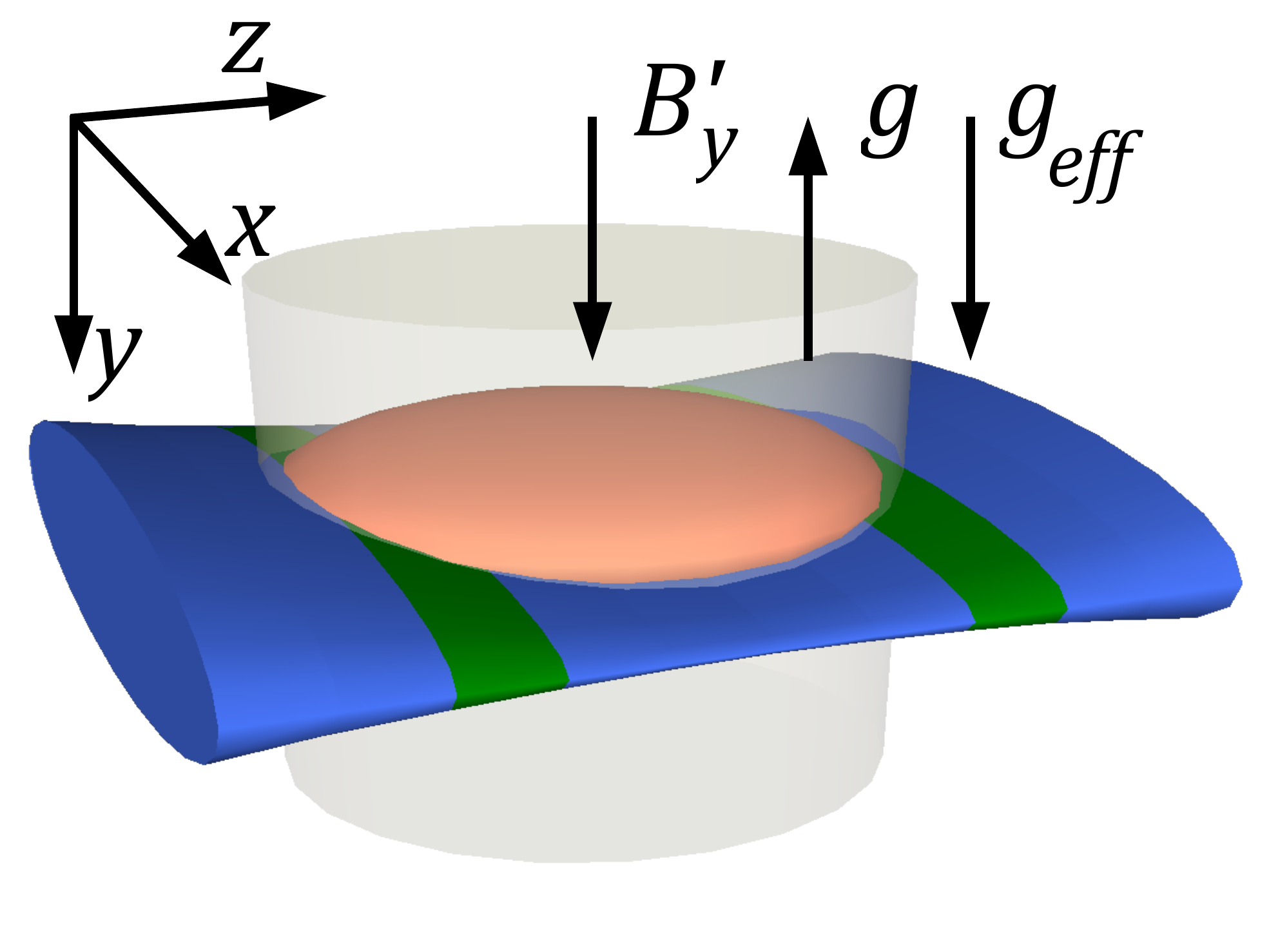}
            \put (2,75) {(a)}
        \end{overpic}\\

        \begin{overpic}[width=\columnwidth]{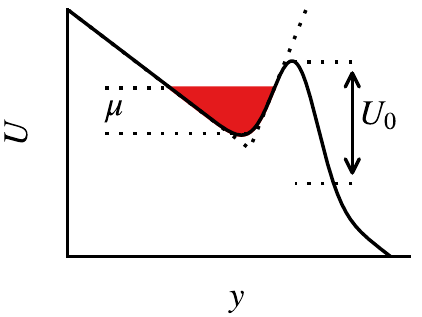}
            \put (2,70) {(c)}
        \end{overpic}
    \end{minipage}%
    \begin{minipage}{0.5\columnwidth}
        \centering
        \begin{overpic}[height=2.53in]{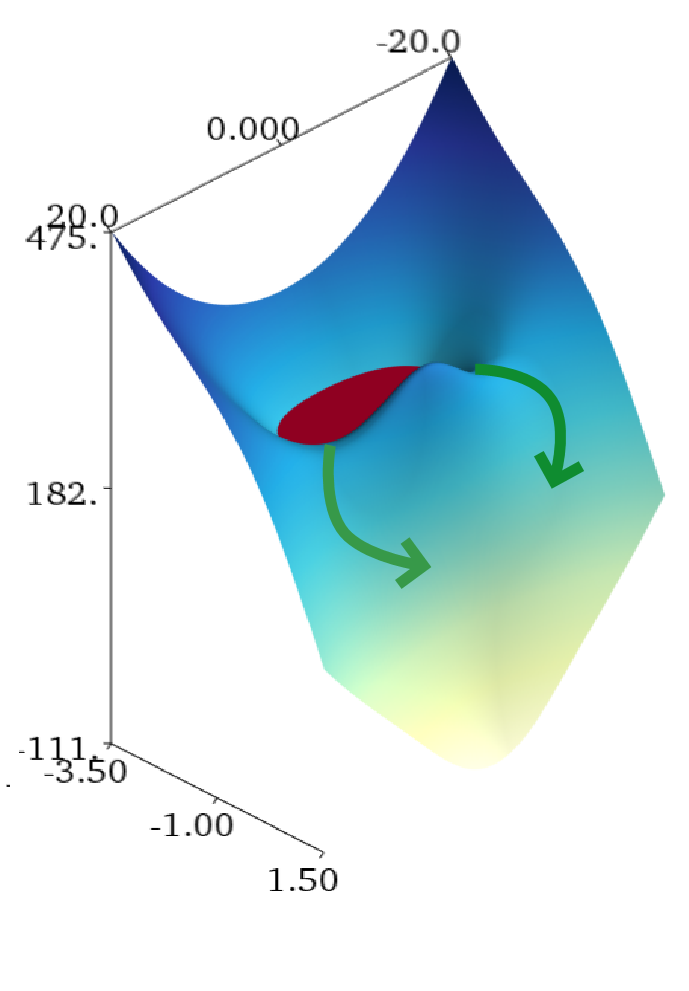}
            \put (-3,100) {(b)}
            \put (15,93) {$z(\si{\micro m})$}
            \put (-5,55) {$U(\si{\nano K})$}
            \put (5,12) {$y(\si{\micro m})$}
        \end{overpic}
    \end{minipage}

\protect\caption{Trap geometry for studying tunneling out of a quasibound state.
(a) A blue-detuned light sheet propagating along the $z$ direction forms one
of the walls of the trap. Anti-Helmholtz coils provide a
vertical field gradient $B'_y$ which over-compensates
gravity, resulting in a net acceleration $g_\mathrm{eff}$ along $+y$
 and harmonic confinement in the $x$ and $z$ directions. The atoms are
trapped in a small pocket created by the light sheet and the field gradient and
tunnel towards $+y$ through two weak links indicated in green.
(b) Surface plot of the potential energy
$U$ at $x=0$. The tight focus of the barrier beam causes it to diffract out,
 increasing the barrier width off-center and
decreasing the barrier height. The initial escape path through the two saddle
points is indicated with green arrows. (c) A slice of the potential energy
at $x=z=0$ shown in solid black. A linear approximation, used to estimate
the chemical potential, is indicated by the dotted lines.
\label{fig:trap_geometry}}
\end{figure}
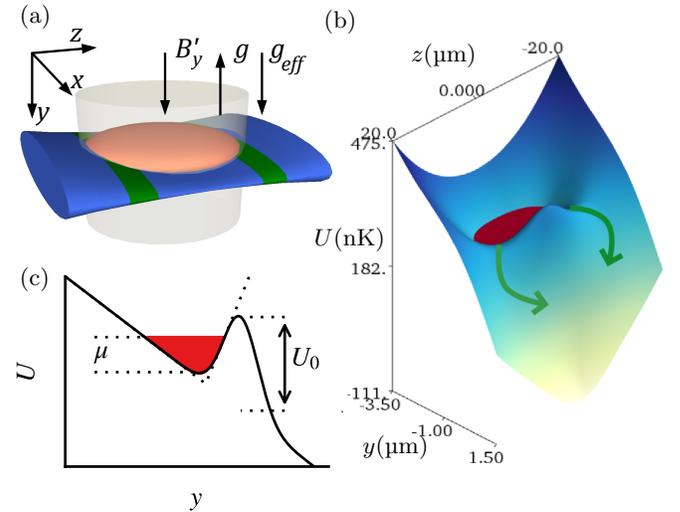

To obtain quasibound states of BECs, we developed a novel trapping geometry in which
a thin repulsive optical barrier forms one of the
walls of the trap, as depicted in Fig.~\ref{fig:trap_geometry}.
The ground state properties of
the condensate in this trap show excellent agreement with mean-field theory.
The strong exponential dependence of the tunneling rate on interactions is demonstrated by
simultaneously measuring the number of atoms left in the trap and inferring the
chemical potential from time-of-flight measurements of the condensate.

Techniques for preparing and manipulating ultra-cold atoms have matured
in the past two decades, leading to a number of fundamental
experiments on quantum tunneling. Experiments studying tunneling
between bound states have explored Josephson oscillations
and self trapping~\cite{albiez2005}, the
DC and AC Josephson effect~\cite{Levy2007},
and the crossover from hydrodynamic and Josephson regimes~\cite{LeBlanc2011}.
In optical lattice systems, the interplay between inter-well tunneling and
strong interactions is seen to give rise to the superfluid to Mott insulator
transition~\cite{Greiner2002,Jordens2008}. Meanwhile, experiments studying tunneling from a
bound state into the continuum are fewer,
and largely study Landau-Zener tunneling out of an optical lattice.
Early work on Landau-Zener tunneling of a BEC demonstrated inter-well coherence in the
tunneling process by observing pulse trains emitted at the Bloch
frequency~\cite{Anderson1998}. Since then, the role of interatomic interactions
has been investigated in Landau-Zener tunneling~\cite{Morsch2001,cristiani2002}.
Deviations from an exponential decay have been demonstrated in Landau-Zener tunneling
outside the BEC context, arising
due to reversible system-environment coupling at early times~\cite{Wilkinson1997},
or due to Zeno and anti-Zeno effects~\cite{Fischer2001}.  However, in our work it is interactions
and the statistical and macroscopic nature of BECs that create a highly non-exponential behavior.

While experiments have focused on tunneling between bound states,
there has been much theoretical activity studying trapped ultra-cold gases tunneling
into the continuum via a thin barrier. The effect of mean-field interactions on the tunneling rate has
been calculated for both attractive and repulsive
interactions, predicting a non-exponential decay curve~\cite{carr2005macroscopic}.
Numerical simulations of dynamics of a trapped condensate tunneling out through a barrier
reveal formations of shock waves inside the condensate, and blips emerging
on the escaped side, as well as the formation of
solitons~\cite{Dekel2010,salasnich2001}. Studies of beyond-mean-field
effects consider tunneling of a strongly interacting Tonks-Girardeau
gas~\cite{DelCampo2006} and the development of correlations and fragmentation
during the tunneling process~\cite{Lode2012}.
Our work opens up the hitherto unexplored experimental regime of
tunneling from a single trapping well into the continuum.

A central feature of our work is a novel Repulsive Sheet Trap (REST),
formed using a combination of a quadrupole magnetic field and a blue detuned
light sheet, as depicted in Fig.~\ref{fig:trap_geometry}.
The magnetic trap provides harmonic confinement in the horizontal directions
with trapping frequencies $\omega_x=2\pi\times\SI{86}{\hertz}$ and
$\omega_z=2\pi\times\SI{43}{\hertz}$. Additionally, it provides a
$\SI{28.3}{G/cm}$ vertical magnetic field gradient, which combined with
gravity results in a net upward acceleration of $g_\mathrm{eff}=\SI{8.4}{m/s^2}$.
A thin light sheet serves as the tunnel barrier and is formed by focusing
a $\SI{405}{\nano\meter}$ laser beam using a high Numerical Aperture (NA)
objective (design based on Ref.~\cite{alt2002}). The beam is nearly Gaussian
with a waist $w_0=\SI{1.3\pm 0.1}{\micro\meter}$ in the $y$ direction and a
Rayleigh range $z_R=\SI{8}{\micro m}$, determined by knife-edge scans.
An acousto-optic deflector is used to scan the beam in the $x$ direction to
create a flat potential within a $\SI{100}{\micro\meter}$
region~\cite{chang2013}.

Our experiment begins with a cloud of $^{87}$Rb atoms in the
$\mid F=2,m_{F}=2\rangle$ ground state evaporatively cooled close to
degeneracy in a hybrid trap~\cite{chang2014}. The magnetic field
gradient is set to cancel gravity in the hybrid trap. The cloud is then
adiabatically transferred to the REST trap by ramping up the barrier
height $U_0$ and the magnetic field
gradient while ramping down the power in the hybrid trap beam. Due to the
small trapping volume of the REST trap, the phase space density increases
during the transfer due to the dimple
effect~\cite{Pinkse1997a,Stamper-Kurn1998,lin2009}. Further evaporation is
achieved by lowering
the barrier height to $\sim\SI{550}{\nano\kelvin}$ to get a pure BEC with 150k
atoms. An RF knife is used transfer the escaped atoms to an untrapped $m_F$ state
and eject them out of the magnetic trap.

To initiate the tunneling dynamics, the barrier height is then non-adiabatically
ramped down in \SI{5}{\milli s}. The condensate is held in the trap for a
variable time from $\SI{0.1}{\milli \second}$ to $\SI{1.2}{\second}$.
The trapping potentials are then abruptly turned off and the cloud is
imaged after a $\SI{20}{\milli \second}$ time-of-flight expansion.
We image on the $F=2 \rightarrow F=3$ cycling transition using $\sigma^+$ light
propagating along $z$ and correct for probe saturation
effects~\cite{Reinaudi2007,egorov2012} to ensure accurate atom number
calibration. The atom number calibration is verified by measuring the
critical temperature in the hybrid trap which agrees within 2\% with the
theoretically predicted value.

\begin{figure}[t]
\includegraphics[width=1\columnwidth]{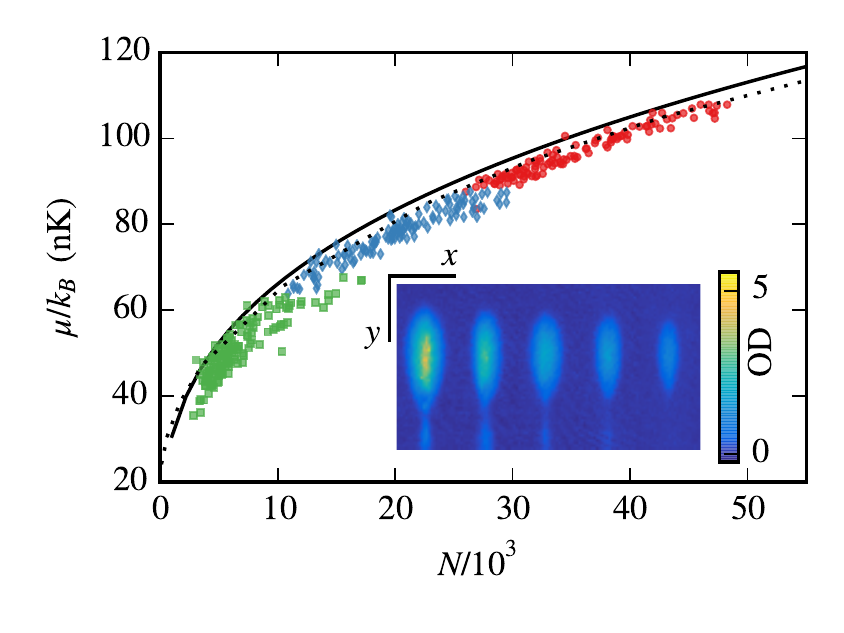}
\protect\caption{The chemical potential $\mu$ of the condensate as a function
of the number of atoms $N$ for barrier heights of \SI{330(35)}{\nano K}
(red circles), \SI{290(30)}{\nano K} (blue diamonds) and
\SI{240(25)}{\nano K} (green squares). Black solid line is an estimate using
Eq.~\ref{eq:mu_infty_wall} and the dotted line is a numerical solution of the
3D Gross-Pitaevskii Equation for a barrier height of \SI{300}{\nano K}.
The inset shows a collage of absorption images with progressively decreasing
atom number from left to right. A trail of atoms espacing along the $y$ direction
is seen when the atom number is high. OD is the on-resonance
optical density of the cloud after correcting for probe saturation
effects.\label{fig:mu_vs_n}}
\end{figure}

The expansion of the condensate during time of flight is highly anisotropic,
as seen in the inset of Fig. \ref{fig:mu_vs_n}. The tight confinement in the
$y$ direction causes the condensate to rapidly expand in the $y$ direction,
converting its interaction energy to kinetic
energy~\cite{Baym1996,PhysRevLett.77.5315}.
We extract the chemical potential from the final $y$ width by fitting the
2D column density to an inverted paraboloid integrated along the imaging axis.
This distribution, while strictly valid only for harmonic traps, fits our data
well. Given our unusual trap geometry, there is no analytical expression for
the chemical potential in the Thomas-Fermi limit. However, we can approximate
the Gaussian barrier with a linear potential
(see Fig.~\hyperref[fig:trap_geometry]{\ref{fig:trap_geometry}(c)})
with an acceleration $a_{b}=2U_{0}/mw_0\sqrt{e}-g_\mathrm{eff}$ to get an analytical
approximation for the chemical potential:
\begin{equation}
\mu=\left\{ 12\left(\hbar\bar{\omega}\right)^{2}\left(m\bar{a}\right)\left(Na_{s}\right)\right\} ^{1/3}.\label{eq:mu_infty_wall}
\end{equation}
Here $\bar{\omega}=\sqrt{\omega_{x}\omega_{z}}$, $N$ is the number of atoms,
$a_s$ is the s-wave scattering length, $m$ is the mass of the particle
and $\bar{a}=g_\mathrm{eff}a_b/(g_\mathrm{eff}+a_b)$ is the reduced acceleration.

Fig.~\ref{fig:mu_vs_n} shows the chemical potential of the condensate for three
different barrier heights. The barrier height $U_0$ is calculated by
measuring the power in the barrier beam and calculating the AC Stark
shift~\cite{grimm2000optical}. The reported uncertainty is due to systematic
errors in estimating the transmitted fraction of the barrier beam through all
the optics and due to uncertainty in the measured barrier waist.
The chemical potential data agrees well with the approximation in
Eq.~\ref{eq:mu_infty_wall}, evaluated using measured trap parameters. However,
due to the tight confinement in the $y$ direction, we are in a regime where
$\mu$ is comparable to the single particle ground state energy
$\epsilon_{0}=\left(\hbar m\bar{a}^{2}/2\right)^{1/3}\sim \SI{20}{\nano K}$.
Thus, kinetic energy corrections to the Thomas-Fermi approximation are important~\cite{Lundh1997,Mateo2007}.
Indeed, the agreement with data is better when the chemical potential is
calculated by numerically solving the full 3D Gross-Pitaevskii Equation (GPE)
equation, as seen in Fig.~\ref{fig:mu_vs_n}.


To characterize loss processes other than escape through the barrier, we study
the decay from a trap with a high barrier height of \SI{700}{\nano K}.
At this barrier height, escape due to classical spilling and tunneling is
negligible. We find that in the \SI{1.5}{\second} observation time, the decay
is exponential with a decay rate of
$\Gamma_\mathrm{bg}=\SI{0.31\pm 0.02}{\per \second}$, which we take as our
background decay rate. This rate is consistent with the three-body recombination
rate, given by $\Gamma_\mathrm{3b}=L\langle n^2\rangle$, where $L$ in the
three-body decay constant. For the REST trap in the Thomas-Fermi limit,
we can show that $\left\langle n^{2}\right\rangle =3n_{0}^{2}/10=\left(3/10\right)\left(\mu/g\right)^{2}$,
where $n_0$ is the peak density. Using the value for $L$ from
Ref.~\cite{Soding2014} and the measured chemical potential $\mu=\SI{92}{\nano K}$,
we find that $\Gamma_\mathrm{3b}=\SI{0.34\pm0.09}{\per\second}$.
We do not find any discernible thermal component emerge during the hold.
Thus, we can ignore losses due to heating and thermal activation.

\begin{figure}[t]
\includegraphics[width=1\columnwidth]{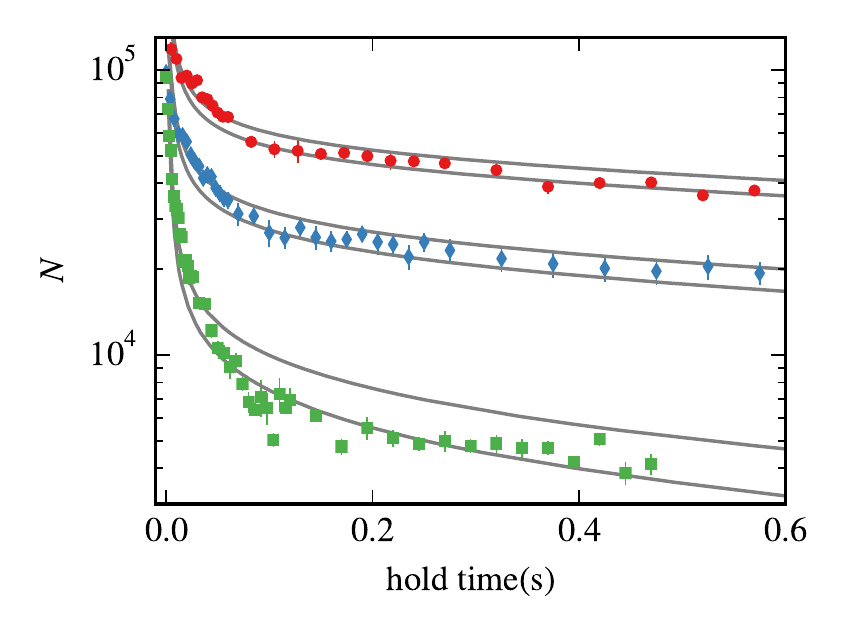}
\protect\caption{Number of atoms $N$ left in the trap after a hold time $t$
for barrier heights of \SI{330(35)}{\nano K} (red circles),
\SI{290(30)}{\nano K} (blue diamonds) and \SI{240(25)}{\nano K}
(green squares). Solid gray lines are results of 3D GPE simulations for a
barrier height of 230, 240, 290, 300, 340 and 350 nK from bottom to top.
Only the first \SI{0.6}{s} are shown here since the decay rate approaches the
background decay rate at later times.\label{fig:n-vs-t}}
\end{figure}

\begin{figure}[t]
\begin{overpic}[width=1\columnwidth]{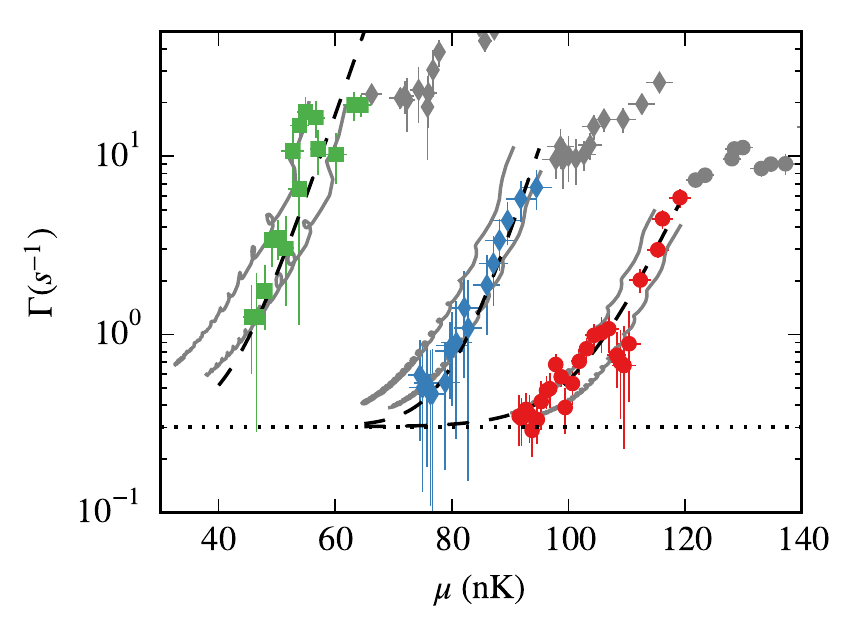}
\put (3, 70) {(a)}
\end{overpic}
\begin{overpic}[width=1\columnwidth]{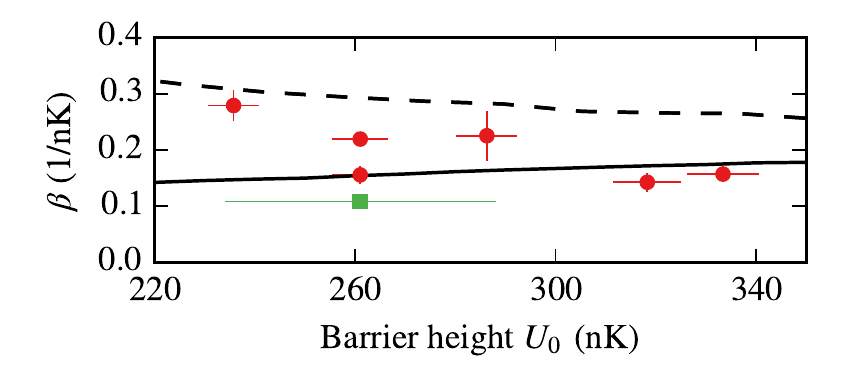}
\put (3, 45) {(b)}
\end{overpic}
\protect\caption{(a) Decay rate $\Gamma$ as a function of the chemical potential
$\mu$. Horizontal dotted line indicates the background decay rate. Data points
where the chemical potential is greater than the trap depth $U_s$ are shown in
gray. The dashed line is a fit to the
function $\Gamma=\Gamma_{\mathrm{bg}}+\exp\left(\alpha+\beta\mu\right)$ and
solid gray lines are results of 3D GPE simulations.
Data points shown in gray are not used for fitting, and correspond to classical spilling
The color schemes match
those shown in Fig.~\ref{fig:n-vs-t}. (b) The slope $\beta$ as a function of
barrier height. Vertical error bars are obtained from the fit. Horizontal error
bars represent statistical fluctuations in the barrier power, except for the
green square, which represents the systematic error.
Solid and dashed lines are results of 3D GPE simulations 1D
single particle transfer matrix calculations respectively.
\label{fig:decay_rate}}
\end{figure}

Next, we discuss the escape dynamics of the condensate when the barrier height
is lowered. Fig.~\ref{fig:n-vs-t} shows the number of atoms remaining in the
trap with time on a semi-log plot. An exponential decay process, characterized
by a constant decay rate, would appear as a straight line on a semi-log plot,
whereas here we see a dramatic decrease in the decay rate with time. The decay
rate $\Gamma=d\ln N/dt$, calculated by fitting sets of
5 consecutive points to a parabola and evaluating the slope at the center,
is shown in Fig.~\hyperref[fig:decay_rate]{\ref{fig:decay_rate}(a)}.

We identify three distinct regimes in the decay process: (a) classical spilling
over the barrier in the first 10-\SI{20}{ms}, (b) quantum tunneling
from \SI{20}{ms} to 0.5\si{s} and (c) decay dominated by background losses
from \SI{500}{ms} onward. The initial non-adiabatic lowering of the barrier
height causes the condensate to rapidly expand and spill over the two saddle
points of the trap, as shown in
Fig~\hyperref[fig:trap_geometry]{\ref{fig:trap_geometry}(b)}. The \SI{5}{ms}
initial ramp-down time of the barrier height is chosen to be comparable to
$1/\omega_z=\SI{3.7}{ms}$, so that it is slow
enough not to cause sloshing in the trap after the ramp down, but fast enough
that tunneling does not begin as the barrier is being ramped down. We expect
the spill to occur on a similar timescale, and indeed we see that in the
first 10-20 ms, $\mu$ drops to below the trap depth $U_s$, at which point the
decay transitions from classical spill to quantum tunneling.

The trap depth $U_s$, which is the difference in potential energy at the saddle
point and the bottom of the trap, is calculated from the peak barrier
height $U_0$ and measured trap parameters. In
Fig.~\hyperref[fig:decay_rate]{\ref{fig:decay_rate}(a)},
the points where the chemical potential is greater than $U_s$ are shown in
gray. Close to the transition point where $\mu \sim U_s$, the $\Gamma$ vs $\mu$
data shows a kink and the decay rate starts dropping exponentially after the
transition. This provides confirmation that the decay mechanism has switched
from classical spilling to tunneling.
The tunneling regime is characterized by
an exponential dependence of the decay rate on the chemical potential.
There is a small range of $\mu$ of about
\SI{20}{\nano K} for which the tunneling range is appreciable. In this range
the decay rate $\Gamma$ drops dramatically until it reaches the background
decay rate.

The experimental results are compared against full 3D GPE simulations, with
measured trap parameters and initial atom numbers used
in the simulations. The ground state is first found by imaginary time
propagation in a trap with a high barrier height. Mimicking
the experiment, the barrier height is then lowered using a linear ramp to a
final value. Absorbing boundary conditions were added to avoid reflections of
the escaped atoms from the edge of the grid. The strength of the absorber
smoothly increases with distance from the barrier to avoid reflections
from the absorber. From Fig.~\ref{fig:n-vs-t} and Fig.~\ref{fig:decay_rate},
we see that our data agrees well with the simulations. Oscillations
in the chemical potential curves in Fig.~\ref{fig:decay_rate} are due to
breathing mode oscillations surviving after the initial spill out.
The small disagreement of our data with simulations could be attributed to a
systematic error in estimating the barrier height and the slight non-Gaussian nature of
the barrier beam due to residual spherical aberrations in the focused beam.
Fig. \hyperref[fig:decay_rate]{\ref{fig:decay_rate}b} compares the slope $\beta$
of the $\log(\Gamma)-\mu$ curves against results results of both 3D GPE simulations
and a 1D transfer matrix calculation with a barrier height $U_s$ and
width equal to the waist size at the saddle point of the potential.
A WKB-style argument suggests an exponential dependence on $\mu$, where the
steepness depends primarily on the barrier thickness. Here we see that
both the single-particle and GPE simulations give exponential behavior,
with slopes between $0.15$ and \SI{0.3}{nK^{-1}}, consistent with our data.

In conclusion, using a novel trapping configuration, we have demonstrated for
the first time quantum tunneling of a condensate from a single trapping well
into the continuum, and shown the exponential dependence of the tunneling rate
on the chemical potential. Having shown good agreement with mean-field simulations,
our trapping geometry may be extended to observe tunneling of many-body
states. For low atom numbers of around
100-1000, $\mu$ would be comparable to $\epsilon_0$, which freezes out all the
dynamics in the vertical ($y$) direction, making the condensate two dimensional.
By further confining the atoms by scanning the barrier beam in both the $x$ and
$y$ directions in a U-shaped pattern, even create one-dimensional condensates could be created.
Tunneling dynamics
of 1D or 2D condensates, where phase fluctuations and defects
have been seen~\cite{Richard2003,Stock2005}, would be an intriguing future direction
of research\cite{Glick2011}. Tunneling out of the REST trap occurs through two symmetric
points, and the tunneled atoms recombine at a time $\pi/2\omega_z$.
The contrast of the resulting interference
pattern could be used as a probe of the coherence of tunneled atoms, and
fragmentation of the condensate~\cite{Lode2012}. Studies of manifestly quantum
phenomenon such as tunneling of macroscopic systems like these can be used to test
the validity of quantum mechanics at the macroscopic level, and investigate
the crossover from the quantum to classical~\cite{Leggett1980}.

\begin{acknowledgements}
The authors thank Rockson Chang, David Spierings, Diego Alcala and Xinxin Zhao
for helpful discussions and A. Stummer for technical support.
Computations were performed on the gpc
supercomputer at the SciNet HPC Consortium~\cite{Loken2010}. SciNet is funded by
the Canada Foundation for Innovation under the auspices of Compute Canada;
the Government of Ontario; Ontario Research Fund - Research Excellence;
and the University of Toronto. Finally, we acknowledge support from NSERC,
CIFAR, NSF, ASOFR and Northrop Grumman Aerospace Systems.
\end{acknowledgements}

\bibliography{refs}

\begin{thebibliography}{38}%
\makeatletter
\providecommand \@ifxundefined [1]{%
 \@ifx{#1\undefined}
}%
\providecommand \@ifnum [1]{%
 \ifnum #1\expandafter \@firstoftwo
 \else \expandafter \@secondoftwo
 \fi
}%
\providecommand \@ifx [1]{%
 \ifx #1\expandafter \@firstoftwo
 \else \expandafter \@secondoftwo
 \fi
}%
\providecommand \natexlab [1]{#1}%
\providecommand \enquote  [1]{``#1''}%
\providecommand \bibnamefont  [1]{#1}%
\providecommand \bibfnamefont [1]{#1}%
\providecommand \citenamefont [1]{#1}%
\providecommand \href@noop [0]{\@secondoftwo}%
\providecommand \href [0]{\begingroup \@sanitize@url \@href}%
\providecommand \@href[1]{\@@startlink{#1}\@@href}%
\providecommand \@@href[1]{\endgroup#1\@@endlink}%
\providecommand \@sanitize@url [0]{\catcode `\\12\catcode `\$12\catcode
  `\&12\catcode `\#12\catcode `\^12\catcode `\_12\catcode `\%12\relax}%
\providecommand \@@startlink[1]{}%
\providecommand \@@endlink[0]{}%
\providecommand \url  [0]{\begingroup\@sanitize@url \@url }%
\providecommand \@url [1]{\endgroup\@href {#1}{\urlprefix }}%
\providecommand \urlprefix  [0]{URL }%
\providecommand \Eprint [0]{\href }%
\providecommand \doibase [0]{http://dx.doi.org/}%
\providecommand \selectlanguage [0]{\@gobble}%
\providecommand \bibinfo  [0]{\@secondoftwo}%
\providecommand \bibfield  [0]{\@secondoftwo}%
\providecommand \translation [1]{[#1]}%
\providecommand \BibitemOpen [0]{}%
\providecommand \bibitemStop [0]{}%
\providecommand \bibitemNoStop [0]{.\EOS\space}%
\providecommand \EOS [0]{\spacefactor3000\relax}%
\providecommand \BibitemShut  [1]{\csname bibitem#1\endcsname}%
\let\auto@bib@innerbib\@empty
\bibitem [{\citenamefont {Gurney}\ and\ \citenamefont
  {Condon}(1929)}]{Gurney1929}%
  \BibitemOpen
  \bibfield  {author} {\bibinfo {author} {\bibfnamefont {R.~W.}\ \bibnamefont
  {Gurney}}\ and\ \bibinfo {author} {\bibfnamefont {E.~U.}\ \bibnamefont
  {Condon}},\ }\href {\doibase 10.1103/PhysRev.33.127} {\bibfield  {journal}
  {\bibinfo  {journal} {Physical Review}\ }\textbf {\bibinfo {volume} {33}},\
  \bibinfo {pages} {127} (\bibinfo {year} {1929})}\BibitemShut {NoStop}%
\bibitem [{\citenamefont {Gatteschi}\ and\ \citenamefont
  {Sessoli}(2003)}]{Gatteschi2003}%
  \BibitemOpen
  \bibfield  {author} {\bibinfo {author} {\bibfnamefont {D.}~\bibnamefont
  {Gatteschi}}\ and\ \bibinfo {author} {\bibfnamefont {R.}~\bibnamefont
  {Sessoli}},\ }\href {\doibase 10.1002/anie.200390099} {\bibfield  {journal}
  {\bibinfo  {journal} {Angewandte Chemie International Edition}\ }\textbf
  {\bibinfo {volume} {42}},\ \bibinfo {pages} {268} (\bibinfo {year}
  {2003})}\BibitemShut {NoStop}%
\bibitem [{\citenamefont {Collini}\ \emph {et~al.}(2010)\citenamefont
  {Collini}, \citenamefont {Wong}, \citenamefont {Wilk}, \citenamefont {Curmi},
  \citenamefont {Brumer},\ and\ \citenamefont {Scholes}}]{Collini2010}%
  \BibitemOpen
  \bibfield  {author} {\bibinfo {author} {\bibfnamefont {E.}~\bibnamefont
  {Collini}}, \bibinfo {author} {\bibfnamefont {C.~Y.}\ \bibnamefont {Wong}},
  \bibinfo {author} {\bibfnamefont {K.~E.}\ \bibnamefont {Wilk}}, \bibinfo
  {author} {\bibfnamefont {P.~M.~G.}\ \bibnamefont {Curmi}}, \bibinfo {author}
  {\bibfnamefont {P.}~\bibnamefont {Brumer}}, \ and\ \bibinfo {author}
  {\bibfnamefont {G.~D.}\ \bibnamefont {Scholes}},\ }\href {\doibase
  10.1038/nature08811} {\bibfield  {journal} {\bibinfo  {journal} {Nature}\
  }\textbf {\bibinfo {volume} {463}},\ \bibinfo {pages} {644} (\bibinfo {year}
  {2010})}\BibitemShut {NoStop}%
\bibitem [{\citenamefont {Choi}\ \emph {et~al.}(2000)\citenamefont {Choi},
  \citenamefont {Ihm}, \citenamefont {Louie},\ and\ \citenamefont
  {Cohen}}]{Choi2000}%
  \BibitemOpen
  \bibfield  {author} {\bibinfo {author} {\bibfnamefont {H.~J.}\ \bibnamefont
  {Choi}}, \bibinfo {author} {\bibfnamefont {J.}~\bibnamefont {Ihm}}, \bibinfo
  {author} {\bibfnamefont {S.~G.}\ \bibnamefont {Louie}}, \ and\ \bibinfo
  {author} {\bibfnamefont {M.~L.}\ \bibnamefont {Cohen}},\ }\href {\doibase
  10.1103/PhysRevLett.84.2917} {\bibfield  {journal} {\bibinfo  {journal}
  {Phys. Rev. Lett.}\ }\textbf {\bibinfo {volume} {84}},\ \bibinfo {pages}
  {2917} (\bibinfo {year} {2000})}\BibitemShut {NoStop}%
\bibitem [{\citenamefont {Albiez}\ \emph {et~al.}(2005)\citenamefont {Albiez},
  \citenamefont {Gati}, \citenamefont {F\"{o}lling}, \citenamefont {Hunsmann},
  \citenamefont {Cristiani},\ and\ \citenamefont {Oberthaler}}]{albiez2005}%
  \BibitemOpen
  \bibfield  {author} {\bibinfo {author} {\bibfnamefont {M.}~\bibnamefont
  {Albiez}}, \bibinfo {author} {\bibfnamefont {R.}~\bibnamefont {Gati}},
  \bibinfo {author} {\bibfnamefont {J.}~\bibnamefont {F\"{o}lling}}, \bibinfo
  {author} {\bibfnamefont {S.}~\bibnamefont {Hunsmann}}, \bibinfo {author}
  {\bibfnamefont {M.}~\bibnamefont {Cristiani}}, \ and\ \bibinfo {author}
  {\bibfnamefont {M.~K.}\ \bibnamefont {Oberthaler}},\ }\href {\doibase
  10.1103/PhysRevLett.95.010402} {\bibfield  {journal} {\bibinfo  {journal}
  {Phys. Rev. Lett.}\ }\textbf {\bibinfo {volume} {95}},\ \bibinfo {pages}
  {10402} (\bibinfo {year} {2005})}\BibitemShut {NoStop}%
\bibitem [{\citenamefont {Levy}\ \emph {et~al.}(2007)\citenamefont {Levy},
  \citenamefont {Lahoud}, \citenamefont {Shomroni},\ and\ \citenamefont
  {Steinhauer}}]{Levy2007}%
  \BibitemOpen
  \bibfield  {author} {\bibinfo {author} {\bibfnamefont {S.}~\bibnamefont
  {Levy}}, \bibinfo {author} {\bibfnamefont {E.}~\bibnamefont {Lahoud}},
  \bibinfo {author} {\bibfnamefont {I.}~\bibnamefont {Shomroni}}, \ and\
  \bibinfo {author} {\bibfnamefont {J.}~\bibnamefont {Steinhauer}},\ }\href
  {\doibase 10.1038/nature06186} {\bibfield  {journal} {\bibinfo  {journal}
  {Nature}\ }\textbf {\bibinfo {volume} {449}},\ \bibinfo {pages} {579}
  (\bibinfo {year} {2007})}\BibitemShut {NoStop}%
\bibitem [{\citenamefont {LeBlanc}\ \emph {et~al.}(2011)\citenamefont
  {LeBlanc}, \citenamefont {Bardon}, \citenamefont {McKeever}, \citenamefont
  {Extavour}, \citenamefont {Jervis}, \citenamefont {Thywissen}, \citenamefont
  {Piazza},\ and\ \citenamefont {Smerzi}}]{LeBlanc2011}%
  \BibitemOpen
  \bibfield  {author} {\bibinfo {author} {\bibfnamefont {L.~J.}\ \bibnamefont
  {LeBlanc}}, \bibinfo {author} {\bibfnamefont {A.~B.}\ \bibnamefont {Bardon}},
  \bibinfo {author} {\bibfnamefont {J.}~\bibnamefont {McKeever}}, \bibinfo
  {author} {\bibfnamefont {M.~H.~T.}\ \bibnamefont {Extavour}}, \bibinfo
  {author} {\bibfnamefont {D.}~\bibnamefont {Jervis}}, \bibinfo {author}
  {\bibfnamefont {J.~H.}\ \bibnamefont {Thywissen}}, \bibinfo {author}
  {\bibfnamefont {F.}~\bibnamefont {Piazza}}, \ and\ \bibinfo {author}
  {\bibfnamefont {A.}~\bibnamefont {Smerzi}},\ }\href {\doibase
  10.1103/PhysRevLett.106.025302} {\bibfield  {journal} {\bibinfo  {journal}
  {Physical Review Letters}\ }\textbf {\bibinfo {volume} {106}},\ \bibinfo
  {pages} {025302} (\bibinfo {year} {2011})}\BibitemShut {NoStop}%
\bibitem [{\citenamefont {Greiner}\ \emph {et~al.}(2002)\citenamefont
  {Greiner}, \citenamefont {Mandel}, \citenamefont {Esslinger}, \citenamefont
  {H\"{a}nsch},\ and\ \citenamefont {Bloch}}]{Greiner2002}%
  \BibitemOpen
  \bibfield  {author} {\bibinfo {author} {\bibfnamefont {M.}~\bibnamefont
  {Greiner}}, \bibinfo {author} {\bibfnamefont {O.}~\bibnamefont {Mandel}},
  \bibinfo {author} {\bibfnamefont {T.}~\bibnamefont {Esslinger}}, \bibinfo
  {author} {\bibfnamefont {T.~W.}\ \bibnamefont {H\"{a}nsch}}, \ and\ \bibinfo
  {author} {\bibfnamefont {I.}~\bibnamefont {Bloch}},\ }\href {\doibase
  10.1038/415039a} {\bibfield  {journal} {\bibinfo  {journal} {Nature}\
  }\textbf {\bibinfo {volume} {415}},\ \bibinfo {pages} {39} (\bibinfo {year}
  {2002})}\BibitemShut {NoStop}%
\bibitem [{\citenamefont {J\"{o}rdens}\ \emph {et~al.}(2008)\citenamefont
  {J\"{o}rdens}, \citenamefont {Strohmaier}, \citenamefont {G\"{u}nter},
  \citenamefont {Moritz},\ and\ \citenamefont {Esslinger}}]{Jordens2008}%
  \BibitemOpen
  \bibfield  {author} {\bibinfo {author} {\bibfnamefont {R.}~\bibnamefont
  {J\"{o}rdens}}, \bibinfo {author} {\bibfnamefont {N.}~\bibnamefont
  {Strohmaier}}, \bibinfo {author} {\bibfnamefont {K.}~\bibnamefont
  {G\"{u}nter}}, \bibinfo {author} {\bibfnamefont {H.}~\bibnamefont {Moritz}},
  \ and\ \bibinfo {author} {\bibfnamefont {T.}~\bibnamefont {Esslinger}},\
  }\href {\doibase 10.1038/nature07244} {\bibfield  {journal} {\bibinfo
  {journal} {Nature}\ }\textbf {\bibinfo {volume} {455}},\ \bibinfo {pages}
  {204} (\bibinfo {year} {2008})}\BibitemShut {NoStop}%
\bibitem [{\citenamefont {Anderson}(1998)}]{Anderson1998}%
  \BibitemOpen
  \bibfield  {author} {\bibinfo {author} {\bibfnamefont {B.~P.}\ \bibnamefont
  {Anderson}},\ }\href {\doibase 10.1126/science.282.5394.1686} {\bibfield
  {journal} {\bibinfo  {journal} {Science}\ }\textbf {\bibinfo {volume}
  {282}},\ \bibinfo {pages} {1686} (\bibinfo {year} {1998})}\BibitemShut
  {NoStop}%
\bibitem [{\citenamefont {Morsch}\ \emph {et~al.}(2001)\citenamefont {Morsch},
  \citenamefont {M\"{u}ller}, \citenamefont {Cristiani}, \citenamefont
  {Ciampini},\ and\ \citenamefont {Arimondo}}]{Morsch2001}%
  \BibitemOpen
  \bibfield  {author} {\bibinfo {author} {\bibfnamefont {O.}~\bibnamefont
  {Morsch}}, \bibinfo {author} {\bibfnamefont {J.~H.}\ \bibnamefont
  {M\"{u}ller}}, \bibinfo {author} {\bibfnamefont {M.}~\bibnamefont
  {Cristiani}}, \bibinfo {author} {\bibfnamefont {D.}~\bibnamefont {Ciampini}},
  \ and\ \bibinfo {author} {\bibfnamefont {E.}~\bibnamefont {Arimondo}},\
  }\href {\doibase 10.1103/PhysRevLett.87.140402} {\bibfield  {journal}
  {\bibinfo  {journal} {Physical Review Letters}\ }\textbf {\bibinfo {volume}
  {87}},\ \bibinfo {pages} {140402} (\bibinfo {year} {2001})}\BibitemShut
  {NoStop}%
\bibitem [{\citenamefont {Cristiani}\ \emph {et~al.}(2002)\citenamefont
  {Cristiani}, \citenamefont {Morsch}, \citenamefont {M\"{u}ller},
  \citenamefont {Ciampini},\ and\ \citenamefont {Arimondo}}]{cristiani2002}%
  \BibitemOpen
  \bibfield  {author} {\bibinfo {author} {\bibfnamefont {M.}~\bibnamefont
  {Cristiani}}, \bibinfo {author} {\bibfnamefont {O.}~\bibnamefont {Morsch}},
  \bibinfo {author} {\bibfnamefont {J.~H.}\ \bibnamefont {M\"{u}ller}},
  \bibinfo {author} {\bibfnamefont {D.}~\bibnamefont {Ciampini}}, \ and\
  \bibinfo {author} {\bibfnamefont {E.}~\bibnamefont {Arimondo}},\ }\href
  {\doibase 10.1103/PhysRevA.65.063612} {\bibfield  {journal} {\bibinfo
  {journal} {Phys. Rev. A}\ }\textbf {\bibinfo {volume} {65}},\ \bibinfo
  {pages} {63612} (\bibinfo {year} {2002})}\BibitemShut {NoStop}%
\bibitem [{\citenamefont {Wilkinson}\ \emph {et~al.}(1997)\citenamefont
  {Wilkinson}, \citenamefont {Bharucha}, \citenamefont {Fischer}, \citenamefont
  {Madison}, \citenamefont {Morrow}, \citenamefont {Niu}, \citenamefont
  {Sundaram},\ and\ \citenamefont {Raizen}}]{Wilkinson1997}%
  \BibitemOpen
  \bibfield  {author} {\bibinfo {author} {\bibfnamefont {S.~R.}\ \bibnamefont
  {Wilkinson}}, \bibinfo {author} {\bibfnamefont {C.~F.}\ \bibnamefont
  {Bharucha}}, \bibinfo {author} {\bibfnamefont {M.~C.}\ \bibnamefont
  {Fischer}}, \bibinfo {author} {\bibfnamefont {K.~W.}\ \bibnamefont
  {Madison}}, \bibinfo {author} {\bibfnamefont {P.~R.}\ \bibnamefont {Morrow}},
  \bibinfo {author} {\bibfnamefont {Q.}~\bibnamefont {Niu}}, \bibinfo {author}
  {\bibfnamefont {B.}~\bibnamefont {Sundaram}}, \ and\ \bibinfo {author}
  {\bibfnamefont {M.~G.}\ \bibnamefont {Raizen}},\ }\href {\doibase
  10.1038/42418} {\ \textbf {\bibinfo {volume} {387}},\ \bibinfo {pages} {575}
  (\bibinfo {year} {1997})}\BibitemShut {NoStop}%
\bibitem [{\citenamefont {Fischer}\ \emph {et~al.}(2001)\citenamefont
  {Fischer}, \citenamefont {Guti{\'{e}}rrez-Medina},\ and\ \citenamefont
  {Raizen}}]{Fischer2001}%
  \BibitemOpen
  \bibfield  {author} {\bibinfo {author} {\bibfnamefont {M.~C.}\ \bibnamefont
  {Fischer}}, \bibinfo {author} {\bibfnamefont {B.}~\bibnamefont
  {Guti{\'{e}}rrez-Medina}}, \ and\ \bibinfo {author} {\bibfnamefont {M.~G.}\
  \bibnamefont {Raizen}},\ }\href {\doibase 10.1103/PhysRevLett.87.040402}
  {\bibfield  {journal} {\bibinfo  {journal} {Physical review letters}\
  }\textbf {\bibinfo {volume} {87}},\ \bibinfo {pages} {040402} (\bibinfo
  {year} {2001})}\BibitemShut {NoStop}%
\bibitem [{\citenamefont {Carr}\ \emph {et~al.}(2005)\citenamefont {Carr},
  \citenamefont {Holland},\ and\ \citenamefont
  {Malomed}}]{carr2005macroscopic}%
  \BibitemOpen
  \bibfield  {author} {\bibinfo {author} {\bibfnamefont {L.~D.}\ \bibnamefont
  {Carr}}, \bibinfo {author} {\bibfnamefont {M.~J.}\ \bibnamefont {Holland}}, \
  and\ \bibinfo {author} {\bibfnamefont {B.~A.}\ \bibnamefont {Malomed}},\
  }\href@noop {} {\bibfield  {journal} {\bibinfo  {journal} {Journal of Physics
  B: Atomic, Molecular and Optical Physics}\ }\textbf {\bibinfo {volume}
  {38}},\ \bibinfo {pages} {3217} (\bibinfo {year} {2005})}\BibitemShut
  {NoStop}%
\bibitem [{\citenamefont {Dekel}\ \emph {et~al.}(2010)\citenamefont {Dekel},
  \citenamefont {Farberovich}, \citenamefont {Fleurov},\ and\ \citenamefont
  {Soffer}}]{Dekel2010}%
  \BibitemOpen
  \bibfield  {author} {\bibinfo {author} {\bibfnamefont {G.}~\bibnamefont
  {Dekel}}, \bibinfo {author} {\bibfnamefont {V.}~\bibnamefont {Farberovich}},
  \bibinfo {author} {\bibfnamefont {V.}~\bibnamefont {Fleurov}}, \ and\
  \bibinfo {author} {\bibfnamefont {A.}~\bibnamefont {Soffer}},\ }\href
  {\doibase 10.1103/PhysRevA.81.063638} {\bibfield  {journal} {\bibinfo
  {journal} {Physical Review A}\ }\textbf {\bibinfo {volume} {81}},\ \bibinfo
  {pages} {063638} (\bibinfo {year} {2010})}\BibitemShut {NoStop}%
\bibitem [{\citenamefont {Salasnich}\ \emph {et~al.}(2001)\citenamefont
  {Salasnich}, \citenamefont {Parola},\ and\ \citenamefont
  {Reatto}}]{salasnich2001}%
  \BibitemOpen
  \bibfield  {author} {\bibinfo {author} {\bibfnamefont {L.}~\bibnamefont
  {Salasnich}}, \bibinfo {author} {\bibfnamefont {A.}~\bibnamefont {Parola}}, \
  and\ \bibinfo {author} {\bibfnamefont {L.}~\bibnamefont {Reatto}},\ }\href
  {\doibase 10.1103/PhysRevA.64.023601} {\bibfield  {journal} {\bibinfo
  {journal} {Phys. Rev. A}\ }\textbf {\bibinfo {volume} {64}},\ \bibinfo
  {pages} {23601} (\bibinfo {year} {2001})}\BibitemShut {NoStop}%
\bibitem [{\citenamefont {del Campo}\ \emph {et~al.}(2006)\citenamefont {del
  Campo}, \citenamefont {Delgado}, \citenamefont {Garc\'{\i}a-Calder\'{o}n},
  \citenamefont {Muga},\ and\ \citenamefont {Raizen}}]{DelCampo2006}%
  \BibitemOpen
  \bibfield  {author} {\bibinfo {author} {\bibfnamefont {A.}~\bibnamefont {del
  Campo}}, \bibinfo {author} {\bibfnamefont {F.}~\bibnamefont {Delgado}},
  \bibinfo {author} {\bibfnamefont {G.}~\bibnamefont
  {Garc\'{\i}a-Calder\'{o}n}}, \bibinfo {author} {\bibfnamefont {J.~G.}\
  \bibnamefont {Muga}}, \ and\ \bibinfo {author} {\bibfnamefont {M.~G.}\
  \bibnamefont {Raizen}},\ }\href {\doibase 10.1103/PhysRevA.74.013605}
  {\bibfield  {journal} {\bibinfo  {journal} {Physical Review A}\ }\textbf
  {\bibinfo {volume} {74}},\ \bibinfo {pages} {013605} (\bibinfo {year}
  {2006})}\BibitemShut {NoStop}%
\bibitem [{\citenamefont {Lode}\ \emph {et~al.}(2012)\citenamefont {Lode},
  \citenamefont {Streltsov}, \citenamefont {Sakmann}, \citenamefont {Alon},\
  and\ \citenamefont {Cederbaum}}]{Lode2012}%
  \BibitemOpen
  \bibfield  {author} {\bibinfo {author} {\bibfnamefont {A.~U.~J.}\
  \bibnamefont {Lode}}, \bibinfo {author} {\bibfnamefont {A.~I.}\ \bibnamefont
  {Streltsov}}, \bibinfo {author} {\bibfnamefont {K.}~\bibnamefont {Sakmann}},
  \bibinfo {author} {\bibfnamefont {O.~E.}\ \bibnamefont {Alon}}, \ and\
  \bibinfo {author} {\bibfnamefont {L.~S.}\ \bibnamefont {Cederbaum}},\ }\href
  {\doibase 10.1073/pnas.1201345109} {\bibfield  {journal} {\bibinfo  {journal}
  {Proceedings of the National Academy of Sciences of the United States of
  America}\ }\textbf {\bibinfo {volume} {109}},\ \bibinfo {pages} {13521}
  (\bibinfo {year} {2012})}\BibitemShut {NoStop}%
\bibitem [{\citenamefont {Alt}(2002)}]{alt2002}%
  \BibitemOpen
  \bibfield  {author} {\bibinfo {author} {\bibfnamefont {W.}~\bibnamefont
  {Alt}},\ }\href {\doibase DOI: 10.1078/0030-4026-00133} {\bibfield  {journal}
  {\bibinfo  {journal} {Optik - International Journal for Light and Electron
  Optics}\ }\textbf {\bibinfo {volume} {113}},\ \bibinfo {pages} {142}
  (\bibinfo {year} {2002})}\BibitemShut {NoStop}%
\bibitem [{\citenamefont {Chang}\ \emph {et~al.}(2013)\citenamefont {Chang},
  \citenamefont {Potnis}, \citenamefont {Ellenor}, \citenamefont {Siercke},
  \citenamefont {Hayat},\ and\ \citenamefont {Steinberg}}]{chang2013}%
  \BibitemOpen
  \bibfield  {author} {\bibinfo {author} {\bibfnamefont {R.}~\bibnamefont
  {Chang}}, \bibinfo {author} {\bibfnamefont {S.}~\bibnamefont {Potnis}},
  \bibinfo {author} {\bibfnamefont {C.~W.}\ \bibnamefont {Ellenor}}, \bibinfo
  {author} {\bibfnamefont {M.}~\bibnamefont {Siercke}}, \bibinfo {author}
  {\bibfnamefont {A.}~\bibnamefont {Hayat}}, \ and\ \bibinfo {author}
  {\bibfnamefont {A.~M.}\ \bibnamefont {Steinberg}},\ }\href {\doibase
  10.1103/PhysRevA.88.053634} {\bibfield  {journal} {\bibinfo  {journal} {Phys.
  Rev. A}\ }\textbf {\bibinfo {volume} {88}},\ \bibinfo {pages} {53634}
  (\bibinfo {year} {2013})}\BibitemShut {NoStop}%
\bibitem [{\citenamefont {Chang}\ \emph {et~al.}(2014)\citenamefont {Chang},
  \citenamefont {Potnis}, \citenamefont {Ramos}, \citenamefont {Zhuang},
  \citenamefont {Hallaji}, \citenamefont {Hayat}, \citenamefont {Duque-Gomez},
  \citenamefont {Sipe},\ and\ \citenamefont {Steinberg}}]{chang2014}%
  \BibitemOpen
  \bibfield  {author} {\bibinfo {author} {\bibfnamefont {R.}~\bibnamefont
  {Chang}}, \bibinfo {author} {\bibfnamefont {S.}~\bibnamefont {Potnis}},
  \bibinfo {author} {\bibfnamefont {R.}~\bibnamefont {Ramos}}, \bibinfo
  {author} {\bibfnamefont {C.}~\bibnamefont {Zhuang}}, \bibinfo {author}
  {\bibfnamefont {M.}~\bibnamefont {Hallaji}}, \bibinfo {author} {\bibfnamefont
  {A.}~\bibnamefont {Hayat}}, \bibinfo {author} {\bibfnamefont
  {F.}~\bibnamefont {Duque-Gomez}}, \bibinfo {author} {\bibfnamefont {J.~E.}\
  \bibnamefont {Sipe}}, \ and\ \bibinfo {author} {\bibfnamefont {A.~M.}\
  \bibnamefont {Steinberg}},\ }\href {\doibase 10.1103/PhysRevLett.112.170404}
  {\bibfield  {journal} {\bibinfo  {journal} {Phys. Rev. Lett.}\ }\textbf
  {\bibinfo {volume} {112}},\ \bibinfo {pages} {170404} (\bibinfo {year}
  {2014})}\BibitemShut {NoStop}%
\bibitem [{\citenamefont {Pinkse}\ \emph {et~al.}(1997)\citenamefont {Pinkse},
  \citenamefont {Mosk}, \citenamefont {Weidem{\"{u}}ller}, \citenamefont
  {Reynolds}, \citenamefont {Hijmans},\ and\ \citenamefont
  {Walraven}}]{Pinkse1997a}%
  \BibitemOpen
  \bibfield  {author} {\bibinfo {author} {\bibfnamefont {P.~W.~H.}\
  \bibnamefont {Pinkse}}, \bibinfo {author} {\bibfnamefont {A.}~\bibnamefont
  {Mosk}}, \bibinfo {author} {\bibfnamefont {M.}~\bibnamefont
  {Weidem{\"{u}}ller}}, \bibinfo {author} {\bibfnamefont {M.~W.}\ \bibnamefont
  {Reynolds}}, \bibinfo {author} {\bibfnamefont {T.~W.}\ \bibnamefont
  {Hijmans}}, \ and\ \bibinfo {author} {\bibfnamefont {J.~T.~M.}\ \bibnamefont
  {Walraven}},\ }\href {\doibase 10.1103/PhysRevLett.78.990} {\bibfield
  {journal} {\bibinfo  {journal} {Physical Review Letters}\ }\textbf {\bibinfo
  {volume} {78}},\ \bibinfo {pages} {990} (\bibinfo {year} {1997})}\BibitemShut
  {NoStop}%
\bibitem [{\citenamefont {Stamper-Kurn}\ \emph {et~al.}(1998)\citenamefont
  {Stamper-Kurn}, \citenamefont {Miesner}, \citenamefont {Chikkatur},
  \citenamefont {Inouye}, \citenamefont {Stenger},\ and\ \citenamefont
  {Ketterle}}]{Stamper-Kurn1998}%
  \BibitemOpen
  \bibfield  {author} {\bibinfo {author} {\bibfnamefont {D.~M.}\ \bibnamefont
  {Stamper-Kurn}}, \bibinfo {author} {\bibfnamefont {H.-J.}\ \bibnamefont
  {Miesner}}, \bibinfo {author} {\bibfnamefont {A.~P.}\ \bibnamefont
  {Chikkatur}}, \bibinfo {author} {\bibfnamefont {S.}~\bibnamefont {Inouye}},
  \bibinfo {author} {\bibfnamefont {J.}~\bibnamefont {Stenger}}, \ and\
  \bibinfo {author} {\bibfnamefont {W.}~\bibnamefont {Ketterle}},\ }\href
  {\doibase 10.1103/PhysRevLett.81.2194} {\bibfield  {journal} {\bibinfo
  {journal} {Physical Review Letters}\ }\textbf {\bibinfo {volume} {81}},\
  \bibinfo {pages} {2194} (\bibinfo {year} {1998})}\BibitemShut {NoStop}%
\bibitem [{\citenamefont {Lin}\ \emph {et~al.}(2009)\citenamefont {Lin},
  \citenamefont {Perry}, \citenamefont {Compton}, \citenamefont {Spielman},\
  and\ \citenamefont {Porto}}]{lin2009}%
  \BibitemOpen
  \bibfield  {author} {\bibinfo {author} {\bibfnamefont {Y.-J.}\ \bibnamefont
  {Lin}}, \bibinfo {author} {\bibfnamefont {A.~R.}\ \bibnamefont {Perry}},
  \bibinfo {author} {\bibfnamefont {R.~L.}\ \bibnamefont {Compton}}, \bibinfo
  {author} {\bibfnamefont {I.~B.}\ \bibnamefont {Spielman}}, \ and\ \bibinfo
  {author} {\bibfnamefont {J.~V.}\ \bibnamefont {Porto}},\ }\href {\doibase
  10.1103/PhysRevA.79.063631} {\bibfield  {journal} {\bibinfo  {journal} {Phys.
  Rev. A}\ }\textbf {\bibinfo {volume} {79}},\ \bibinfo {pages} {63631}
  (\bibinfo {year} {2009})}\BibitemShut {NoStop}%
\bibitem [{\citenamefont {Reinaudi}\ \emph {et~al.}(2007)\citenamefont
  {Reinaudi}, \citenamefont {Lahaye}, \citenamefont {Wang},\ and\ \citenamefont
  {Gu\'{e}ry-Odelin}}]{Reinaudi2007}%
  \BibitemOpen
  \bibfield  {author} {\bibinfo {author} {\bibfnamefont {G.}~\bibnamefont
  {Reinaudi}}, \bibinfo {author} {\bibfnamefont {T.}~\bibnamefont {Lahaye}},
  \bibinfo {author} {\bibfnamefont {Z.}~\bibnamefont {Wang}}, \ and\ \bibinfo
  {author} {\bibfnamefont {D.}~\bibnamefont {Gu\'{e}ry-Odelin}},\ }\href
  {\doibase 10.1364/OL.32.003143} {\bibfield  {journal} {\bibinfo  {journal}
  {Optics Letters}\ }\textbf {\bibinfo {volume} {32}},\ \bibinfo {pages} {3143}
  (\bibinfo {year} {2007})}\BibitemShut {NoStop}%
\bibitem [{\citenamefont {Egorov}(2012)}]{egorov2012}%
  \BibitemOpen
  \bibfield  {author} {\bibinfo {author} {\bibfnamefont {M.}~\bibnamefont
  {Egorov}},\ }\emph {\bibinfo {title} {Coherence and collective oscillations
  of a two-component Bose-Einstein condensate}},\ \href@noop {} {Ph.D.
  thesis},\ \bibinfo  {school} {Centre for Atom Optics and Ultrafast
  Spectroscopy Swinburne University of Technology, Melbourne, Australia.}
  (\bibinfo {year} {2012})\BibitemShut {NoStop}%
\bibitem [{\citenamefont {Baym}\ and\ \citenamefont
  {Pethick}(1996)}]{Baym1996}%
  \BibitemOpen
  \bibfield  {author} {\bibinfo {author} {\bibfnamefont {G.}~\bibnamefont
  {Baym}}\ and\ \bibinfo {author} {\bibfnamefont {C.~J.}\ \bibnamefont
  {Pethick}},\ }\href {\doibase 10.1103/PhysRevLett.76.6} {\bibfield  {journal}
  {\bibinfo  {journal} {Physical Review Letters}\ }\textbf {\bibinfo {volume}
  {76}},\ \bibinfo {pages} {6} (\bibinfo {year} {1996})}\BibitemShut {NoStop}%
\bibitem [{\citenamefont {Castin}\ and\ \citenamefont
  {Dum}(1996)}]{PhysRevLett.77.5315}%
  \BibitemOpen
  \bibfield  {author} {\bibinfo {author} {\bibfnamefont {Y.}~\bibnamefont
  {Castin}}\ and\ \bibinfo {author} {\bibfnamefont {R.}~\bibnamefont {Dum}},\
  }\href {\doibase 10.1103/PhysRevLett.77.5315} {\bibfield  {journal} {\bibinfo
   {journal} {Phys. Rev. Lett.}\ }\textbf {\bibinfo {volume} {77}},\ \bibinfo
  {pages} {5315} (\bibinfo {year} {1996})}\BibitemShut {NoStop}%
\bibitem [{\citenamefont {Grimm}\ \emph {et~al.}(2000)\citenamefont {Grimm},
  \citenamefont {Weidem\"{u}ller},\ and\ \citenamefont
  {Ovchinnikov}}]{grimm2000optical}%
  \BibitemOpen
  \bibfield  {author} {\bibinfo {author} {\bibfnamefont {R.}~\bibnamefont
  {Grimm}}, \bibinfo {author} {\bibfnamefont {M.}~\bibnamefont
  {Weidem\"{u}ller}}, \ and\ \bibinfo {author} {\bibfnamefont {Y.~B.}\
  \bibnamefont {Ovchinnikov}},\ }\href@noop {} {\bibfield  {journal} {\bibinfo
  {journal} {Advances in atomic, molecular, and optical physics}\ }\textbf
  {\bibinfo {volume} {42}},\ \bibinfo {pages} {95} (\bibinfo {year}
  {2000})}\BibitemShut {NoStop}%
\bibitem [{\citenamefont {Lundh}\ \emph {et~al.}(1997)\citenamefont {Lundh},
  \citenamefont {Pethick},\ and\ \citenamefont {Smith}}]{Lundh1997}%
  \BibitemOpen
  \bibfield  {author} {\bibinfo {author} {\bibfnamefont {E.}~\bibnamefont
  {Lundh}}, \bibinfo {author} {\bibfnamefont {C.~J.}\ \bibnamefont {Pethick}},
  \ and\ \bibinfo {author} {\bibfnamefont {H.}~\bibnamefont {Smith}},\ }\href
  {\doibase 10.1103/PhysRevA.55.2126} {\bibfield  {journal} {\bibinfo
  {journal} {Physical Review A}\ }\textbf {\bibinfo {volume} {55}},\ \bibinfo
  {pages} {2126} (\bibinfo {year} {1997})}\BibitemShut {NoStop}%
\bibitem [{\citenamefont {Mateo}\ and\ \citenamefont
  {Delgado}(2007)}]{Mateo2007}%
  \BibitemOpen
  \bibfield  {author} {\bibinfo {author} {\bibfnamefont {A.~M.~n.}\
  \bibnamefont {Mateo}}\ and\ \bibinfo {author} {\bibfnamefont
  {V.}~\bibnamefont {Delgado}},\ }\href {\doibase 10.1103/PhysRevA.75.063610}
  {\bibfield  {journal} {\bibinfo  {journal} {Physical Review A}\ }\textbf
  {\bibinfo {volume} {75}},\ \bibinfo {pages} {063610} (\bibinfo {year}
  {2007})}\BibitemShut {NoStop}%
\bibitem [{\citenamefont {S\"{o}ding}\ \emph {et~al.}(2014)\citenamefont
  {S\"{o}ding}, \citenamefont {Gu\'{e}ry-Odelin}, \citenamefont {Desbiolles},
  \citenamefont {Chevy}, \citenamefont {Inamori},\ and\ \citenamefont
  {Dalibard}}]{Soding2014}%
  \BibitemOpen
  \bibfield  {author} {\bibinfo {author} {\bibfnamefont {J.}~\bibnamefont
  {S\"{o}ding}}, \bibinfo {author} {\bibfnamefont {D.}~\bibnamefont
  {Gu\'{e}ry-Odelin}}, \bibinfo {author} {\bibfnamefont {P.}~\bibnamefont
  {Desbiolles}}, \bibinfo {author} {\bibfnamefont {F.}~\bibnamefont {Chevy}},
  \bibinfo {author} {\bibfnamefont {H.}~\bibnamefont {Inamori}}, \ and\
  \bibinfo {author} {\bibfnamefont {J.}~\bibnamefont {Dalibard}},\ }\href
  {\doibase 10.1007/s003400050805} {\bibfield  {journal} {\bibinfo  {journal}
  {Applied Physics B}\ }\textbf {\bibinfo {volume} {69}},\ \bibinfo {pages}
  {257} (\bibinfo {year} {2014})}\BibitemShut {NoStop}%
\bibitem [{\citenamefont {Richard}\ \emph {et~al.}(2003)\citenamefont
  {Richard}, \citenamefont {Gerbier}, \citenamefont {Thywissen}, \citenamefont
  {Hugbart}, \citenamefont {Bouyer},\ and\ \citenamefont
  {Aspect}}]{Richard2003}%
  \BibitemOpen
  \bibfield  {author} {\bibinfo {author} {\bibfnamefont {S.}~\bibnamefont
  {Richard}}, \bibinfo {author} {\bibfnamefont {F.}~\bibnamefont {Gerbier}},
  \bibinfo {author} {\bibfnamefont {J.~H.}\ \bibnamefont {Thywissen}}, \bibinfo
  {author} {\bibfnamefont {M.}~\bibnamefont {Hugbart}}, \bibinfo {author}
  {\bibfnamefont {P.}~\bibnamefont {Bouyer}}, \ and\ \bibinfo {author}
  {\bibfnamefont {A.}~\bibnamefont {Aspect}},\ }\href {\doibase
  10.1103/PhysRevLett.91.010405} {\bibfield  {journal} {\bibinfo  {journal}
  {Physical Review Letters}\ }\textbf {\bibinfo {volume} {91}},\ \bibinfo
  {pages} {010405} (\bibinfo {year} {2003})}\BibitemShut {NoStop}%
\bibitem [{\citenamefont {Stock}\ \emph {et~al.}(2005)\citenamefont {Stock},
  \citenamefont {Hadzibabic}, \citenamefont {Battelier}, \citenamefont
  {Cheneau},\ and\ \citenamefont {Dalibard}}]{Stock2005}%
  \BibitemOpen
  \bibfield  {author} {\bibinfo {author} {\bibfnamefont {S.}~\bibnamefont
  {Stock}}, \bibinfo {author} {\bibfnamefont {Z.}~\bibnamefont {Hadzibabic}},
  \bibinfo {author} {\bibfnamefont {B.}~\bibnamefont {Battelier}}, \bibinfo
  {author} {\bibfnamefont {M.}~\bibnamefont {Cheneau}}, \ and\ \bibinfo
  {author} {\bibfnamefont {J.}~\bibnamefont {Dalibard}},\ }\href {\doibase
  10.1103/PhysRevLett.95.190403} {\bibfield  {journal} {\bibinfo  {journal}
  {Physical Review Letters}\ }\textbf {\bibinfo {volume} {95}},\ \bibinfo
  {pages} {190403} (\bibinfo {year} {2005})}\BibitemShut {NoStop}%
\bibitem [{\citenamefont {Glick}\ and\ \citenamefont {Carr}()}]{Glick2011}%
  \BibitemOpen
  \bibfield  {author} {\bibinfo {author} {\bibfnamefont {J.~A.}\ \bibnamefont
  {Glick}}\ and\ \bibinfo {author} {\bibfnamefont {L.~D.}\ \bibnamefont
  {Carr}},\ }\href {http://arxiv.org/abs/1105.5164} {\ }\Eprint
  {http://arxiv.org/abs/1105.5164} {arXiv:1105.5164} \BibitemShut {NoStop}%
\bibitem [{\citenamefont {Leggett}(1980)}]{Leggett1980}%
  \BibitemOpen
  \bibfield  {author} {\bibinfo {author} {\bibfnamefont {A.~J.}\ \bibnamefont
  {Leggett}},\ }\href {\doibase 10.1143/PTPS.69.80} {\bibfield  {journal}
  {\bibinfo  {journal} {Progress of Theoretical Physics Supplement}\ }\textbf
  {\bibinfo {volume} {69}},\ \bibinfo {pages} {80} (\bibinfo {year}
  {1980})}\BibitemShut {NoStop}%
\bibitem [{\citenamefont {Loken}\ \emph {et~al.}(2010)\citenamefont {Loken},
  \citenamefont {Gruner}, \citenamefont {Groer}, \citenamefont {Peltier},
  \citenamefont {Bunn}, \citenamefont {Craig}, \citenamefont {Henriques},
  \citenamefont {Dempsey}, \citenamefont {Yu}, \citenamefont {Chen},
  \citenamefont {Dursi}, \citenamefont {Chong}, \citenamefont {Northrup},
  \citenamefont {Pinto}, \citenamefont {Knecht},\ and\ \citenamefont
  {Zon}}]{Loken2010}%
  \BibitemOpen
  \bibfield  {author} {\bibinfo {author} {\bibfnamefont {C.}~\bibnamefont
  {Loken}}, \bibinfo {author} {\bibfnamefont {D.}~\bibnamefont {Gruner}},
  \bibinfo {author} {\bibfnamefont {L.}~\bibnamefont {Groer}}, \bibinfo
  {author} {\bibfnamefont {R.}~\bibnamefont {Peltier}}, \bibinfo {author}
  {\bibfnamefont {N.}~\bibnamefont {Bunn}}, \bibinfo {author} {\bibfnamefont
  {M.}~\bibnamefont {Craig}}, \bibinfo {author} {\bibfnamefont
  {T.}~\bibnamefont {Henriques}}, \bibinfo {author} {\bibfnamefont
  {J.}~\bibnamefont {Dempsey}}, \bibinfo {author} {\bibfnamefont {C.-H.}\
  \bibnamefont {Yu}}, \bibinfo {author} {\bibfnamefont {J.}~\bibnamefont
  {Chen}}, \bibinfo {author} {\bibfnamefont {L.~J.}\ \bibnamefont {Dursi}},
  \bibinfo {author} {\bibfnamefont {J.}~\bibnamefont {Chong}}, \bibinfo
  {author} {\bibfnamefont {S.}~\bibnamefont {Northrup}}, \bibinfo {author}
  {\bibfnamefont {J.}~\bibnamefont {Pinto}}, \bibinfo {author} {\bibfnamefont
  {N.}~\bibnamefont {Knecht}}, \ and\ \bibinfo {author} {\bibfnamefont {R.~V.}\
  \bibnamefont {Zon}},\ }\href {\doibase 10.1088/1742-6596/256/1/012026}
  {\bibfield  {journal} {\bibinfo  {journal} {Journal of Physics: Conference
  Series}\ }\textbf {\bibinfo {volume} {256}},\ \bibinfo {pages} {012026}
  (\bibinfo {year} {2010})}\BibitemShut {NoStop}%
\end{thebibliography}%

\end{document}